\documentclass[journal,a4paper]{IEEEtran}
\ifCLASSINFOpdf
\else
\fi
\usepackage{graphicx}
\usepackage{amsmath}
\usepackage{subfigure}
\usepackage{cite}
\usepackage{comment}

\begin{document}

\title{Advanced channel coding for space mission telecommand links}

\author{{M. Baldi, M. Bianchi, F. Chiaraluce, R. Garello, I. Aguilar Sanchez, S. Cioni}

\thanks{Copyright (c) 2013 IEEE. Personal use of this material is permitted. However, permission to use this material for any other purposes must be obtained from the IEEE by sending a request to pubs-permissions@ieee.org.

This work was supported in part by the ESA Contract No 4000106268: Advanced Coding Schemes for Direct Sequence Spread Spectrum Telecommand Links. 
M. Baldi, M. Bianchi and F. Chiaraluce are with Universit\`a Politecnica delle Marche, Dipartimento di Ingegneria dell'Informazione, Ancona, Italy. R. Garello is with Politecnico di Torino, Dipartimento di Elettronica e Telecomunicazioni, Torino, Italy. I. Aguilar Sanchez and S. Cioni are with ESA-ESTEC, TEC-ETC, Noordwijk, The Netherlands.}}
\maketitle
\thispagestyle{empty}

\begin{abstract}
We investigate and compare different options for updating the error correcting code currently used in space mission telecommand links. Taking as a reference the solutions recently emerged as the most promising ones, based on Low-Density Parity-Check codes, we explore the behavior of alternative schemes, based on parallel concatenated turbo codes and soft-decision decoded BCH codes. Our analysis shows that these further options can offer similar or even better performance.
\end{abstract}


\IEEEpeerreviewmaketitle

\section{Introduction}

\IEEEPARstart{T}he only error correcting code currently included in the CCSDS \cite{CCSDS2010} recommendation and the ECSS \cite{ECSS2008} standard for Telecommand (TC) synchronization and channel coding is the expurgated BCH(63, 56) code, with hard-decision decoding. In order to improve such an ``obsolete'' scheme, a lot of work has been recently done to propose solutions that take into account the most recent progress and comply with increasing demand for more and more sophisticated uplink coding capabilities \cite{Calzolari2007}. Among the most attractive options, a prominent role is played by short Low-Density Parity-Check (LDPC) codes, proposed in binary \cite{CCSDS2012} and non-binary \cite{Costantini2012} form.

In this paper, we enlarge the grid of possible candidates to replace the code of the current standard. In particular, we consider parallel turbo codes (already included in the CCSDS Telemetry (TM) recommendation \cite{CCSDS2011}) and extended BCH codes (eBCH) that however, contrary to the standard, use soft-decision decoding in place of hard-decision decoding. As a meaningful example, we focus on the case of codes with length $128$ and rate $1/2$. The performance of the new options are evaluated, also in comparison with the asymptotic limits achievable, i.e., Shannon's sphere packing lower bound (SPLB).
Assuming non-binary LDPC codes as a valuable practical reference,
we show that:
i) new designed parallel turbo codes have performance close to non-binary LDPC codes down to codeword error rate (CER) in the order of $10^{-4}$; ii) the eBCH(128, 64) code decoded through a maximum likelihood (ML) algorithm has nearly optimal performance, better than non-binary LDPC codes. We then focus on sub-optimal algorithms that allow to approach the ML performance, at a reasonable computational cost.

The organization of the paper is as follows. In Section \ref{sec:two} we remind the current standard and the possibility to improve performance by resorting to soft-decision decoding. In Section \ref{sec:three}, we present LDPC codes, that are at the basis of recent proposals for updating the standard. In Section \ref{sec:four} we design parallel turbo codes that are competitive against LDPC codes. In Section \ref{sec:five} we discuss the possibility to apply sub-optimal soft-decision decoding algorithms to eBCH codes, achieving performance very close to that of ML decoding. Finally, some conclusions are drawn in Section \ref{sec:six}, where we also highlight some open issues.

\section{Current standard}
\label{sec:two}

Let us refer to the CCSDS recommendation \cite{CCSDS2010}: it specifies the functions performed in the ``Synchronization and Channel Coding sublayer'' in TC ground-to-space (or space-to-space) communication links. In short, the sublayer takes transfer frames produced by the upper sublayer (``Data Link Protocol sublayer''), elaborates them
and outputs Communications Link Transmission Units (CLTUs) that are passed to the lower layer (``Physical layer'') where they are mapped into the transmitted waveform by
adopting a proper modulation format.
Within the Synchronization and Channel Coding sublayer, three functions are realized: randomization (optional for CCSDS, mandatory for ECSS), error control coding and synchronization.

The current CCSDS recommendation and ECSS standard use a BCH(63, 56) code for error protection against noise and interference. At the receiver side, hard decision is taken on the received symbols.
The performance of the hard-decision decoded BCH(63, 56) code, evaluated on the additive white Gaussian noise channel, are quite unsatisfactory. Because of its very limited error correction capability, it requires very large signal-to-noise ratios. For this code, however, it is possible to perform an effective ML soft-decision decoding based on its trellis representation. More precisely, for each linear code $C(n,k)$ it is possible to apply, for example, the technique described in \cite{Bahl1974}
to build a time-variant trellis representation with a maximum number of states equal to $2^x$, where $x = \min\{k, n - k\}$. For the BCH(63, 56) the maximum number of states is equal to $2^7 = 128$, and then it is possible to apply a soft-decision decoding by using the Viterbi algorithm or the BCJR algorithm \cite{Bahl1974} that, for $128$ states, still have reasonable complexity.

\begin{figure}[tb]
\begin{centering}
\includegraphics[width=83mm,keepaspectratio]{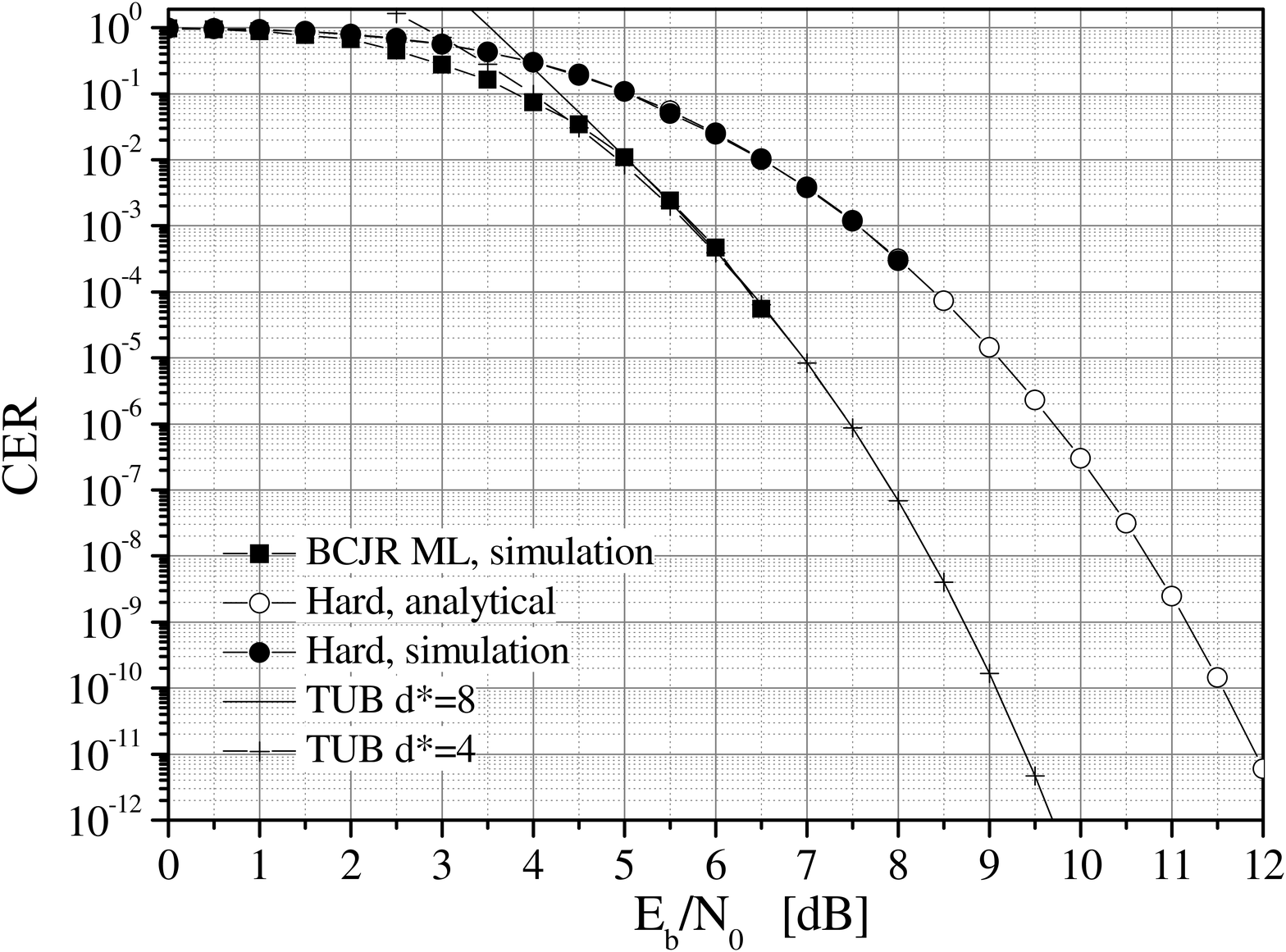}
\caption{Performance of the BCH(63, 56) code with hard- and soft-decision decoding. \label{fig:BCH}}
\par\end{centering}
\end{figure}

In Fig. \ref{fig:BCH} we have reported the simulated CER of the BCH(63, 56) code when using BCJR, in comparison with the performance achieved through hard-decision decoding. The latter can be also determined analytically. As a further benchmark, we have also plotted the so-called ``truncated union bound'' (TUB), which is given by the following expression:
\begin{equation}
CER_{TUB} = \sum_{i=1}^{d^*} \frac{1}{2}A_i {\rm erfc} \sqrt{i \frac{k}{n} \frac{E_b}{N_0}},
\label{eq:TUB}
\end{equation}
where $A_i$ is the weight-$i$ codeword multiplicity and $E_b/N_0$ is the signal-to-noise ratio per bit. The TUB represents an approximation of the complete union bound (for which $d^* = n$) and, as shown in the figure, provides an excellent approximation of the ML decoding performance in the region of significant CER ($\leq 10^{-2}$). Although the whole code distance profile is known for the considered code, the figure shows that $d^* = 8$ (or even $d^* = 4$) is enough to obtain an excellent approximation. Looking at the figure, we can conclude that, at low error rates, the BCJR ML soft-decision decoding provides a gain of more than 2 dB with respect to the hard-decision decoding curve. Though appreciable, such gain is not enough for the expectations of the updated standard. Thus, other solutions must be explored, for example of the type discussed in the next sections.

\section{LDPC codes}
\label{sec:three}

\begin{figure}[tb]
\begin{centering}
\includegraphics[width=83mm,keepaspectratio]{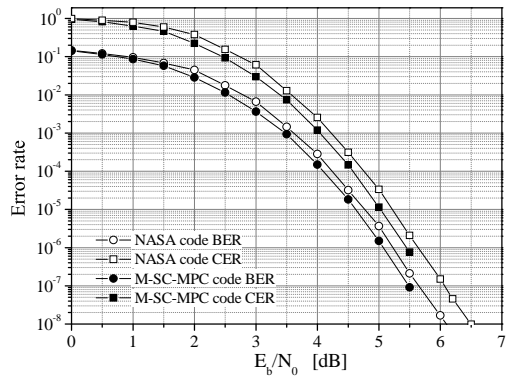}
\caption{Performance of different binary (128, 64) LDPC codes. \label{fig:BinaryLDPC}}
\par\end{centering}
\end{figure}

An obvious way to improve the error rate performance, with respect to the current standard, consists in using more powerful codes, with lower rate \cite{deCola2011}. A first significant proposal, in this sense, has been advanced by the National Aeronautics and Space Administration (NASA) and is described in \cite{CCSDS2012}. It is based on the adoption of three systematic short binary LDPC codes, with rate $1/2$ and length $n = 128, 256$ and $512$, respectively. These codes are designed using protographs with circulant matrices. We have verified that different constructions \cite{Baldi2011a, Baldi2012c, Baldi2012b, Baldi2009a} can be used as well, providing similar performance, with no significant impact on the encoding/decoding complexity. As an example, we have adopted the structure named Multiple Serially-Concatenated Multiple-Parity-Check (M-SC-MPC) code \cite{Baldi2009a}. These codes are a class of structured LDPC codes obtained from the serial concatenation of very simple component codes, named MPC codes, which results in LDPC codes with good performance and very good flexibility in the design. The very simple structure of the component codes also facilitates the encoder implementation. The bit error rate (BER) and CER performance of the $(n = 128, k = 64)$ code designed this way are reported in Fig. \ref{fig:BinaryLDPC} and compared with the performance of the NASA code. Both codes have been decoded by using the sum-product algorithm with log-likelihood ratios (LLR-SPA) (actually, for its code, NASA uses an optimized min* decoding algorithm that has, basically, the same performance) with a maximum number of iterations $I_{\max} = 100$.

Despite the structural differences, Fig. \ref{fig:BinaryLDPC} confirms that the performances of the two codes are very similar. To decide about their goodness, however, an absolute reference is required. Indeed, a valuable benchmark can be provided by the SPLB. Giving a lower bound on the CER performance of a coding scheme with a given codeword length, it is useful to estimate a code ``optimality'', i.e., how far the performance of the considered code is from the best theoretical one, and how much gain is available for other coding schemes, if able to outperform it. Among the various approaches available, the most suitable one is the so-called SP59, as introduced by Shannon in 1959 \cite{Shannon1959}. It must be said that a modified version of this bound is also available (called SP67 \cite{Shannon1967}), that is able to take into account the constraint put by the signal constellation (2-PSK in the present analysis). This further bound has been even improved more recently \cite{Wiechman2008} but such improvements are significant only for high code rates or long codeword lengths, and these conditions are not satisfied by the codes here of interest. Thus, in the present study, we consider SP59 as the most significant SPLB.

\begin{figure}[tb]
\begin{centering}
\includegraphics[width=83mm,keepaspectratio]{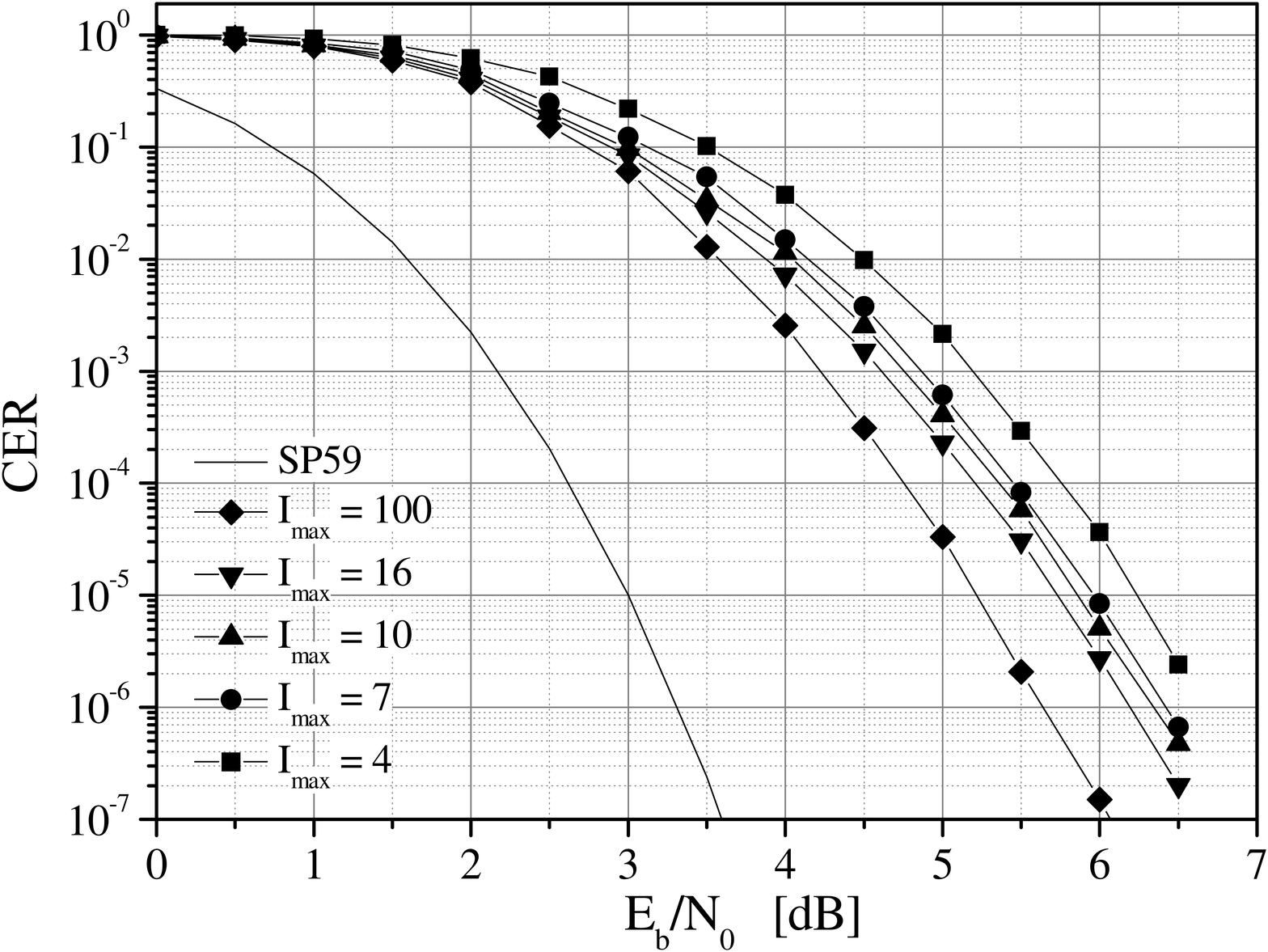}
\caption{Sphere packing lower bound against performance of the NASA (128, 64) LDPC code. \label{fig:SPB}} 
\par\end{centering}
\end{figure}

\begin{figure}[tb]
\begin{centering}
\includegraphics[width=83mm,keepaspectratio]{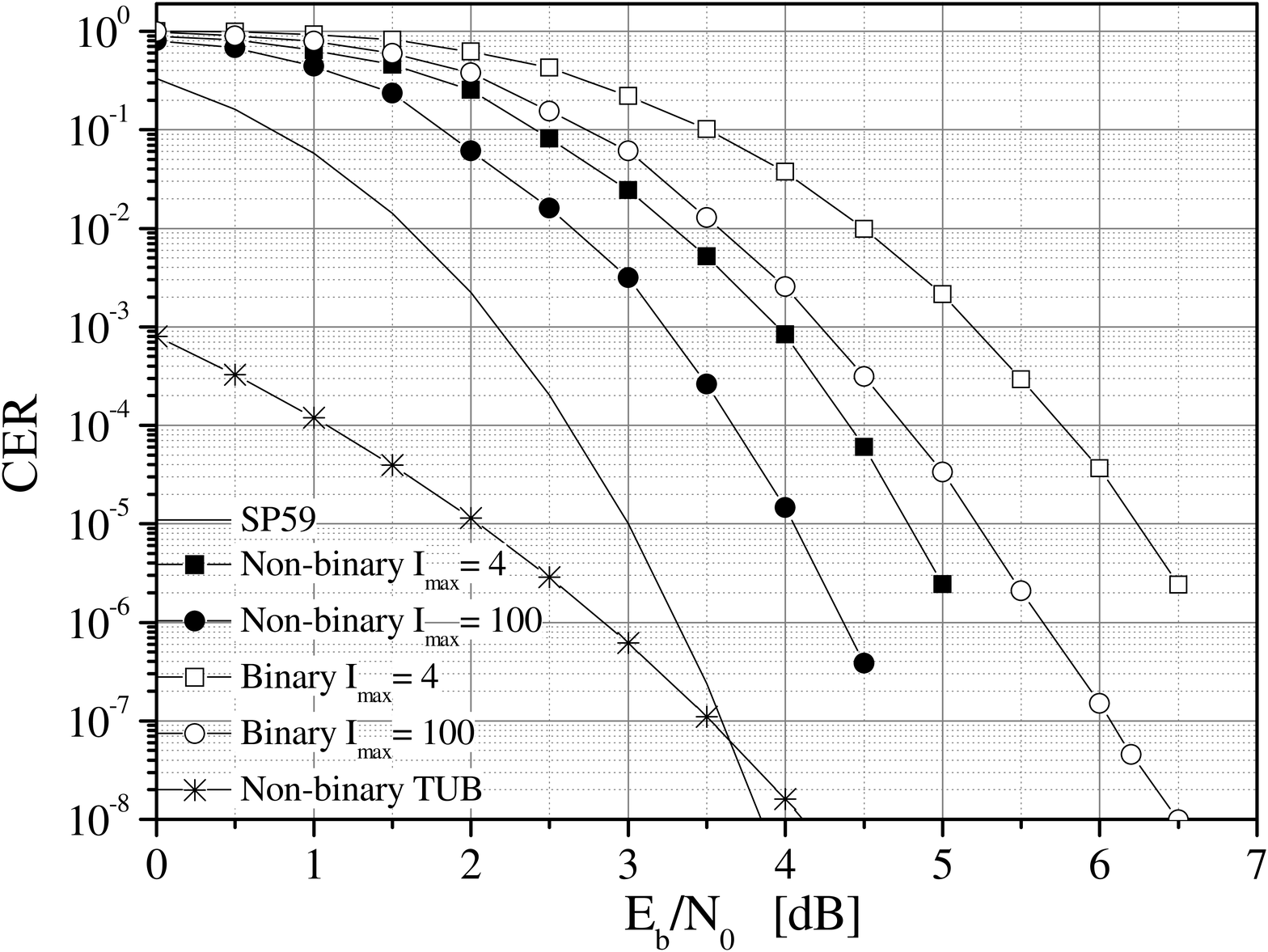}
\caption{Performance of the non-binary (128, 64) LDPC code in comparison with the binary code, the SPLB and the truncated union bound. \label{fig:NonBinaryLDPC}}
\par\end{centering}
\end{figure}

The SP59 is plotted in Fig. \ref{fig:SPB}, for the considered case of $n = 128$ and $k = 64$, and there compared with the performance of the NASA code, for different values of $I_{\max}$. From the figure, we  observe that the performance of the NASA LDPC code is good but not excellent. In fact, its distance from the SPLB is larger than $2$ dB. 
This result suggests that further improvement is potentially achievable. An attractive solution, in such a perspective, consists in using non-binary LDPC codes. These codes have been analyzed in \cite{Costantini2012} and \cite{NASA/JPL2012}. We refer to the implementation in \cite{Costantini2012}, and in Fig. \ref{fig:NonBinaryLDPC} we report the performance of a non-binary LDPC code with $n = 128$ and $k = 64$, constructed on the Galois Field GF(256). Decoding is realized by using iterative algorithms based on fast Hadamard transforms. The TUB has been also plotted, as a further reference, in the figure, since the codeword multiplicity of the non-binary LDPC code is known \cite{Liva2012} (with reference to (\ref{eq:TUB}), $d^* = 14$ is enough for a good representation). However, we note that, being below the SPLB, it is not particularly significant in the explored region. From Fig. \ref{fig:NonBinaryLDPC}, the improvement achievable by using the non-binary code is evident and the distance from the SPLB becomes very small (in the order of $1$ dB in the region of low CER). Hence, these results look excellent. In the next sections, however, we will show that they can be approached and, in principle, even outperformed by using different solutions.

\section{Parallel turbo codes}
\label{sec:four}

Parallel turbo codes (PTCs) are one of the coding options of the CCSDS recommendation for TM links \cite{CCSDS2011}. The CCSDS turbo encoder is based on the parallel concatenation of two equal 16-state systematic convolutional encoders with polynomial description  $(1, (1+D^2 + D^4 + D^5)/(1 + D^3 + D^4))$. The interleavers are based on an algorithmic rule proposed by Berrou and described in \cite{CCSDS2011}. The CCSDS turbo encoder has four possible information frame lengths: $1784, 3568, 7136$ and $8920$ bits. The nominal code rate can be $1/2, 1/3, 1/4$, and $1/6$, but higher rates are obtainable by puncturing.

Maintaining unchanged the encoder structure, we have considered frame lengths shorter than those in the TM recommendation and fixed the nominal code rate to $1/2$, in such a way as to comply with the NASA's choices discussed in the previous section. Because of the shorter length, we cannot use the interleavers in \cite{CCSDS2011} and we must design new smaller interleavers. Among a large number of different options, we have focused attention on: completely random, spread \cite{Divsalar2005}, Quadratic Permutation Polynomial (QPP) \cite{Sun2005} and Dithered Relative Prime (DRP) \cite{Croizer2001} interleavers.

Moreover, since the constituent CCSDS convolutional codes have $16$ states, four extra-tail bits are needed for termination; then, the turbo codeword length is $n = 2(k + 4)$. For the specific case of $k = 64$, this implies to have $n = 136$ and an actual code rate $0.471$. In order to achieve the same code rate of the other schemes (that is necessary for fair comparison), we have implemented a suitable puncturing strategy. More precisely, the algorithm \cite{Garello2001} has been applied to identify the codewords with smallest weight.
Then, we have selected the positions that, with higher probability, do not correspond to bits equal to $1$. Finally we have looked for puncturing patterns insisting on these positions. The design criterion was the maximization of the punctured turbo code minimum distance and the minimization of its multiplicity. 
When no good puncturing patterns were found by this method, we have performed a joint search for both interleaver permutation and puncturing pattern looking for the best punctured turbo codes. In doing this we have adopted periodic puncturing patterns, according to the rules described in \cite{Ryan1999}.

\begin{figure}[tb]
\begin{centering}
\includegraphics[width=93mm,keepaspectratio]{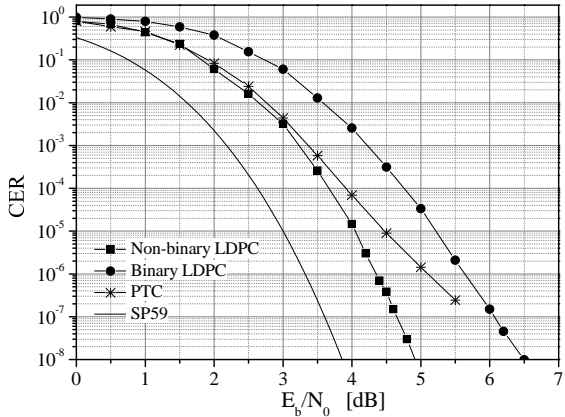}
\caption{Performance of the (128, 64) PTC against binary and non-binary LDPC codes; the SPLB is reported as a reference. \label{fig:PTC}}
\par\end{centering}
\end{figure}

As a result of this optimization process, the best interleaver we have found, among the considered classes, is a DRP interleaver. Using it, the (128, 64) PTC is characterized by minimum distance $d_{\min} = 10$ and weight-$d_{\min}$ codeword multiplicity $A_{\min} = 5$. In Fig. \ref{fig:PTC} we compare the performance of the (128, 64) PTC equipped with such an interleaver with that of the (128, 64) binary LDPC code and the (128, 64) non-binary LDPC code discussed in the previous section. The SP59 is also plotted for the sake of reference. From the figure we see that the performance of the turbo code is very close to that of the non-binary LDPC codes if the requested error rates are not too low. As an example, at CER $\approx 10^{-4}$ the loss is about $0.25$ dB, and becomes about $0.45$ dB at CER $\approx 10^{-5}$. The loss becomes greater for lower and lower CER, because of the higher error floor, due to the smaller minimum distance. 
For the sake of completeness, it must be said that we have developed a similar comparison for the longer codes (that is, with $k = 128$ and $k = 256$). The corresponding CER curves, not reported here because of lack of space, show that the loss for these codes is smaller. Taking this into account, PTCs seem a valid alternative to non-binary LDPC codes for TC applications, at not too low CER values. Besides the error rate performance, the choice of the former or the latter solution may depend on complexity issues, whose evaluation is in progress and will be presented in a next paper.

\section{eBCH codes}
\label{sec:five}
\begin{figure}[tb]
\begin{centering}
\includegraphics[width=93mm,keepaspectratio]{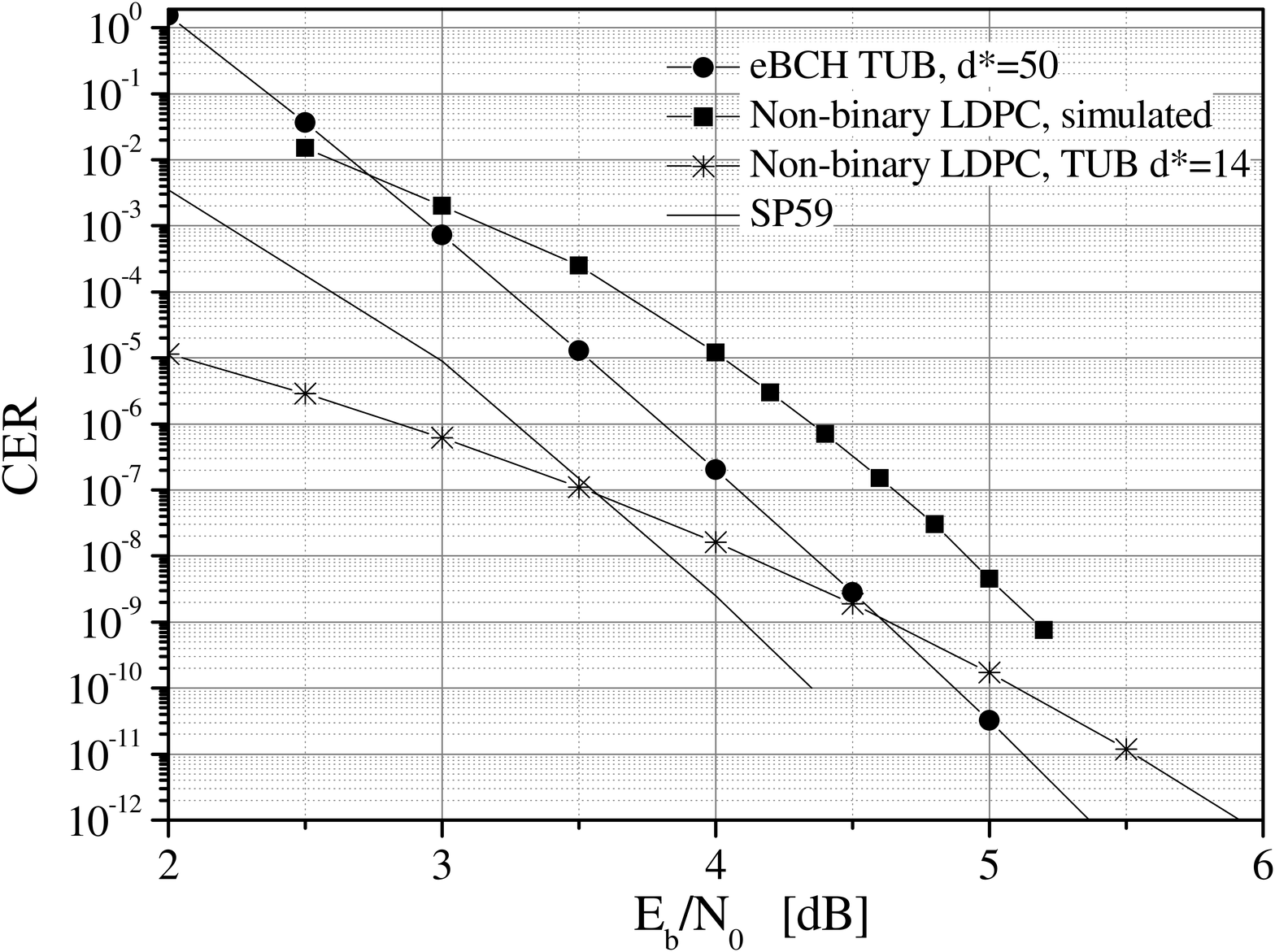}
\caption{TUB of the eBCH(128, 64) code, compared with the TUB and the simulated performance of the non-binary (128, 64) LDPC code and the SPLB. \label{fig:eBCH}}
\par\end{centering}
\end{figure}

\begin{figure}[tb]
\begin{centering}
\includegraphics[width=93mm,keepaspectratio]{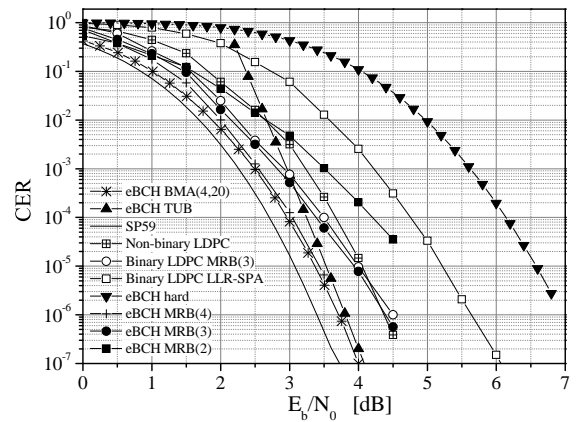}
\caption{Performance of different sub-optimal algorithms for soft-decision decoding of the eBCH(128, 64), in comparison with that of LDPC codes, TUB and SP59. \label{fig:subopt}}
\par\end{centering}
\end{figure}

Within the family of BCH codes, we have considered the eBCH(128, 64) code. The TUB for this code can be easily determined, since its codeword multiplicity is completely known \cite{Morelos2002}.
In particular, it is possible to verify that $d^* = 50$ is enough to have a good description of the 
complete union bound and to obtain a good estimate of an optimal ML soft-decision decoding algorithm, in the region of low error rates. The TUB of the eBCH is shown in Fig. \ref{fig:eBCH} and there compared with the TUB and the simulated performance of the non-binary (128, 64) LDPC code described in Section \ref{sec:three}, as well as with the SP59. From the figure, we see that the gap between the eBCH TUB and the SPLB is very small (for example, $\approx 0.5$ dB for CER $= 10^{-5}$). Moreover, the eBCH TUB achieves a gain of about $0.5$ dB with respect to the simulated non-binary LDPC code with the same length and rate. In explicit terms, this means that if one is able to apply ML decoding to the eBCH(128, 64) code, this can provide performance better than that of all the other solutions discussed so far.
Contrary to the BCH(63, 56) considered in Section \ref{sec:two}, soft-decision decoding of the eBCH(128, 64) code based on its trellis representation is unfeasible. In fact, its trellis has a maximum complexity of $2^{64}$ states. As a consequence, sub-optimal soft-decision decoding algorithms must be applied.
Many sub-optimal algorithms have been presented in the literature.
Most of them are based on ordered statistics decoding. As a further option, one can take advantage of proper LDPC-like code representations \cite{SoftCOM2007}. Among the huge amount of variants available, we have focused attention on some solutions that, in our opinion, are particularly promising, as they permit to conciliate the desire for good performance with the need to maintain limited complexity. More precisely, we have considered:
\begin{itemize}
\item The Box and Match Algorithm (BMA) \cite{Valembois2004a}.
\item The Most Reliable Basis (MRB) algorithm \cite{Wu2007}.
\end{itemize}
Details of these methods can be found in the quoted references and are here omitted, for the sake of brevity. Numerical examples are given in Fig. \ref{fig:subopt}. The BMA curve has been taken from \cite{Valembois2004a}, while all the others have been simulated. For the sake of comparison, the figure also reports:
\begin{itemize}
\item The hard-decision decoding performance of the eBCH code.
\item The TUB of the eBCH code for $d^* = 50$.
\item The SP59 for a (128, 64) code.
\item The performance of the non-binary (128, 64) LDPC code, taken from Section \ref{sec:three}.
\item The performance of the NASA binary (128, 64) LDPC code, taken from Section \ref{sec:two}, and the result of its sub-optimal decoding by using the MRB algorithm.
\end{itemize}
It should be noted that this comparison extends a preliminary analysis, of the same type, previously presented in \cite{deCola2011}.

A number of interesting conclusions can be drawn from the figure. First of all, we observe that the performance of the BMA(4, 20) algorithm (the meaning of the parameters is explained in \cite{Valembois2004a}) is very close to that of the optimal ML soft-decision decoder:
its gap from the SP59 is smaller than $0.5$ dB. Thus, its performance is excellent.
The performance of the MRB(4) (where 4 is the order of the algorithm, see \cite{Wu2007} for details) is practically coincident with that of the BMA(4, 20). On the other hand, tolerating a slight performance degradation, the complexity can be reduced by using the MRB(3): the penalty is limited and the algorithm provides better performance than the non-binary LDPC code down to CER $= 2 \cdot 10^{-6}$. In general, the complexity of the sub-optimal algorithms depends on a number of design parameters \cite{Wu2007} that need to be optimized. However, it is not difficult to find a set of parameters that allow efficient decoding of the eBCH(128, 64) code. Also relevant in the figure, we observe that the MRB(3) algorithm applied to the binary LDPC code yields a significant improvement with respect to the LLR-SPA, and the achieved performance is practically coincident with that of the eBCH code. Taking into account that the complexity of the MRB algorithm is almost independent of the code structure, being only a function of the code parameters, this result confirms the convenience of the eBCH solution over the LDPC one.

\section{Conclusion and open issues}
\label{sec:six}

This paper shows that valid alternatives to the solutions based on binary and non-binary LDPC codes can be found for updating the current TC recommendation. Parallel turbo codes and eBCH codes can provide similar or even better features. More precisely, the turbo code can show a penalty with respect to the non-binary LDPC code but, as a counterpart, it exploits a scheme that is already included in the CCSDS Recommendations and, most of all, it does not suffer some problems related to the possible adoption of the eBCH code. The latter exhibits the best error rate performance. However, extending the sub-optimal decoding algorithms used for the eBCH(128, 64) code to longer codes, while maintaining acceptable complexity, may be difficult.
Additionally, the sub-optimal algorithms generally define ``complete'' decoders.
As well known, this may be a penalty for the undetected frame error rate (UFER) that, in TC applications, is at least as important as the CER (or the frame error rate, FER).
This problem does not exist if a Cyclic Redundancy Check (CRC) code is used for detecting frame integrity, which makes UFER negligible. If the CRC code is not used, the UFER performance can be improved by making the decoder slightly incomplete but this has, obviously, an impact on the CER (and the FER) performance. Another important issue concerns CLTU termination that in the current standard (whose hard-decision decoder is certainly incomplete) is realized by introducing (at the transmitter) and searching for (at the receiver) an uncorrectable pattern. Since such a strategy cannot be applied with complete decoders, different approaches shall be identified for delimiting the CLTU and exploiting the error correction capabilities of the eBCH code.

\section*{Acknowledgment}
The authors wish to thank Marco Chiani and Enrico Paolini for having made available the software to simulate non-binary LDPC codes.

\newcommand{\BIBdecl}{\setlength{\itemsep}{0.01\baselineskip}}
\bibliographystyle{IEEEtran}
\bibliography{Archive}

\end{document}